\documentclass{cmmse2016}
\usepackage{amssymb}
\usepackage[pdftex]{graphicx}

\begin{document}



\newcommand{\Eqref}[1]{(\ref{#1})}


\markboth
  {Quantum Wigner molecules and symmetries} 
  {Yannouleas and Landman}


\title{Quantum Wigner molecules in semiconductor quantum dots and cold-atom optical traps
and their mathematical symmetries} 



\author{\underline{Constantine Yannouleas}}{Constantine.Yannouleas@physics.gateh.edu}{1}

\author{Uzi Landman}{Uzi.Landman@physics.gateh.edu}{1}


\affiliation{1}{School of Physics}
  {Georgia Institute of Technology, Atlanta, GA 30332-0430, USA}


\begin{abstract}
Strong repelling interactions between a few fermions or bosons confined in two-dimensional circular traps 
lead to particle localization and formation of quantum Wigner molecules (QWMs) possessing definite 
point-group space symmetries. These point-group symmetries are "hidden" (or emergent), namely they cannot 
be traced in the circular single-particle densities (SPDs) associated with the exact many-body wave 
functions, but they are manifested as characteristic signatures in the ro-vibrational spectra. An example,
among many, are the few-body QWM states under a high magnetic field or at fast rotation, which are 
precursor states for the fractional quantum Hall effect. The hidden geometric symmetries can be directly 
revealed by using spin-resolved conditional probability distributions, which are extracted from 
configuration-interaction (CI), exact-diagonalization wave functions. The hidden symmetries can also be 
revealed in the CI SPDs by reducing the symmetry of the trap (from circular to elliptic to quasi-linear). 
In addition the hidden symmetries are directly connected to the explicitly broken-symmetry (BS) solutions 
of mean-field approaches, such as unrestricted Hartree-Fock (UHF). A companion step of restoration
of the broken symmetries via projection operators applied on the BS-UHF solutions produces
wave functions directly comparable to the CI ones, and sheds further light into the role played 
by the emergence of hidden symmetries in the exact many-body wave functions. Illustrative examples of the 
importance of hidden symmetries in the many-body problem of few electrons in semiconductor 
quantum dots and of few ultracold atoms in optical traps (where unprecedented control of the interparticle 
interaction has been experimentally achieved recently) will be presented. 

\keywords Wigner molecule, emergent point-group symmetries, broken symmetries, symmetry restoration, 
projection operator, unrestricted Hartree Fock, configuration interaction, 2D semicoductor 
quantum dots, trapped ultracold atoms, fermions, bosons
\end{abstract}
%
%

\begin{figure}[t]
\centering\includegraphics[width=5.2cm]{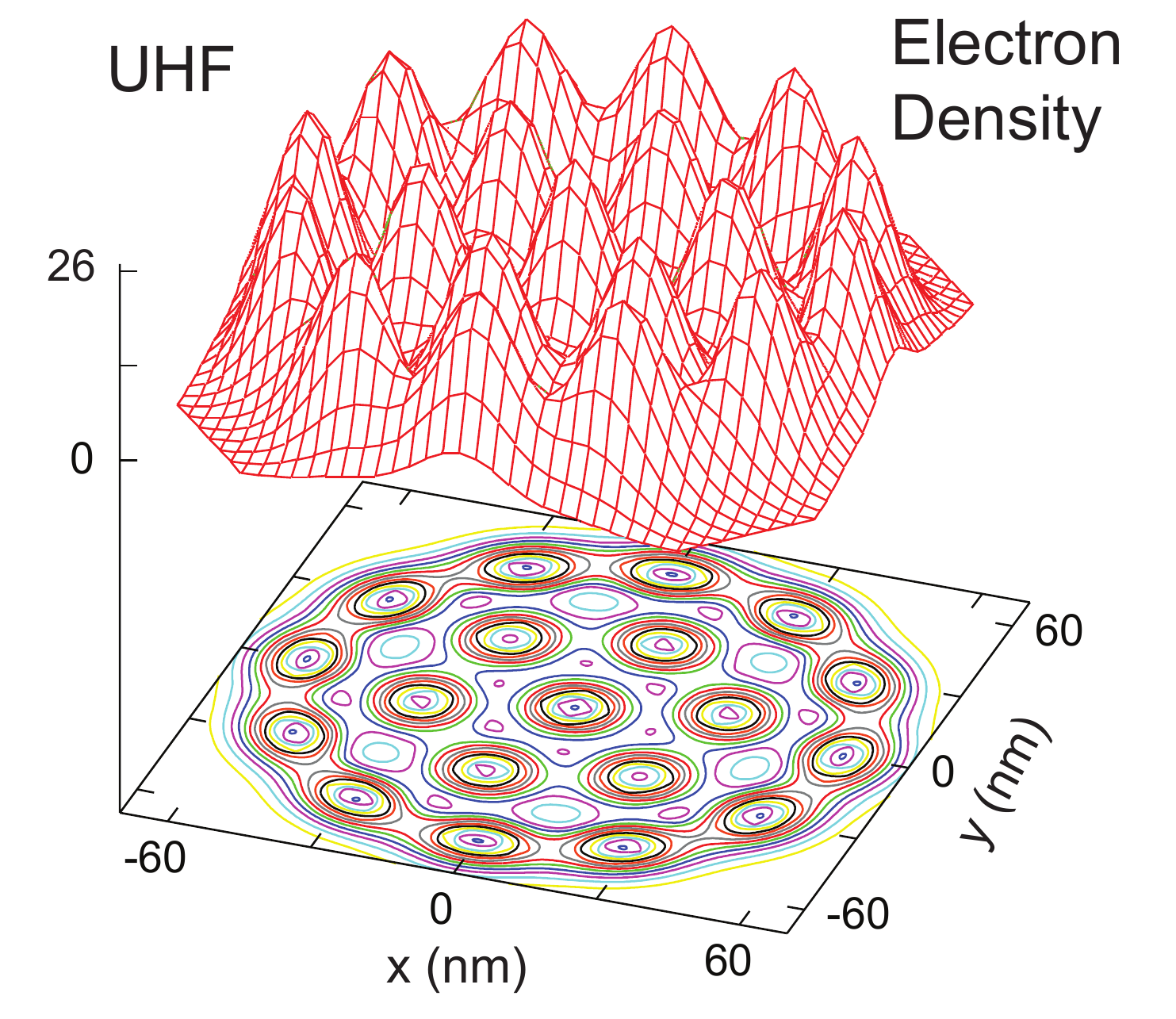}
\caption{
Unrestricted Hartree-Fock electron density in a 2D parabolic QD for $N=19$ 
electrons and total-spin projection $S_z=19/2$, exhibiting
breaking of the circular symmetry at $R_W=5$ and zero magnetic field.
The electrons are (partially) localized in a (1,6,12) multi-ring structure. 
which exhibits point-group symmetries.
Remaining parameters are: parabolic confinement, $\hbar \omega_0=5$ meV;
effective mass $m^*=0.067 m_e$. Distances are in nanometers and the electron 
density in $10^{-4}$ nm$^{-2}$.
}
\end{figure}

\section{Introduction}
This talk focuses on novel, somewhat exotic, types of clusters of few 
fermions or bosons. In particular, we discuss clusters of electrons in manmade 
(artificial) quantum dots (QDs) created through lithographic and gate-voltage 
techniques at semiconductor interfaces, and clusters of neutral ultracold atoms
(either bosonic or fermionic) in harmonic optical traps. These cluster systems 
exhibit interesting emergent physical behavior arising from spontaneous breaking 
of spatial and/or spin symmetries at the {\it mean-field\/} level of theoretical 
treatment \cite{yann99,yann07}; symmetry breaking (SB) is defined 
as a circumstance where a lower energy solution of the Schr\"{o}dinger equation is found 
that is characterized by a lower symmetry than that of the full many-body 
Hamiltonian of the few-body system. Such SB in circular traps directly reflects the localization of 
of particles in cluster arrangements exhibiting point-group symmetries instead of the continuous 
rotational symmetry expected from the many-body Hamiltonian \cite{yann07}. A prominent example is 
the formation of finite electron crystallites (referred to as {\it semi-classical\/} Wigner 
molecules, SCWMs) in two-dimensional (2D) QDs  (see Fig.\ 1). Symmetry breaking at the mean-field 
level is also manifested in the transition \cite{roma04,roma06}, 
induced by increasing the interatomic repulsive contact-interaction strength, of the ground state of
neutral atoms in a parabolic or toroidal 2D trap to a rotating bosonic quantum Wigner molecule (QWM). 
An example is presented in Fig.\ 2, where the hierarchy of the successive approximations (broken symmetry 
UHF $\rightarrow $ symmetry restoration) is illustrated, leading to the symmetry-restored, fully-quantal 
wave function (QWM) in Fig.\ 2 (c,d); the single-particle density (SPD) for the intermediate 
BS-UHF (SCWM) wave function is plotted in Fig.\ 2(b). Note that the point-group symmetry is not visible 
in the SPD after the step of symmetry restoration is executed, i.e., it becomes hidden [see Fig.\ 2(c)], 
but it is revealed via a conditional probability distribution (CPD); see Fig.\ 2(d). 

The CPD gives the probability of finding a particle with spin $\sigma$ at position ${\bf r}$ given that 
another one (referred to as the fixed particle) with spin $\sigma_0$ is located at ${\bf r}_0$.
The degree of particle localization is controlled by the Wigner parameter that 
specifies the strength of the interparticle repulsion relative to the zero-point kinetic energy, i.e.,
$R_W=Z^2e^2/(\kappa l_0 \hbar \omega_0)$ \cite{yann07,roma04} for a Coulomb repulsion 
and $R_\delta=gm/(2\pi\hbar^2)$ \cite{yann07,roma04,roma06} 
for a contact-potential (Dirac-delta) repulsion; $Z$ is the charge of the particle, $\kappa$ is the 
dielectric constant, $\omega_0$ is the frequency of the harmonic 
trap, $l_0=\sqrt{\hbar/(m \omega_0)}$ is the oscillator length, $m$ is the particle mass, and $g$ is the 
strength of the contact interaction.

\begin{figure}[t]
\centering\includegraphics[width=6.50cm]{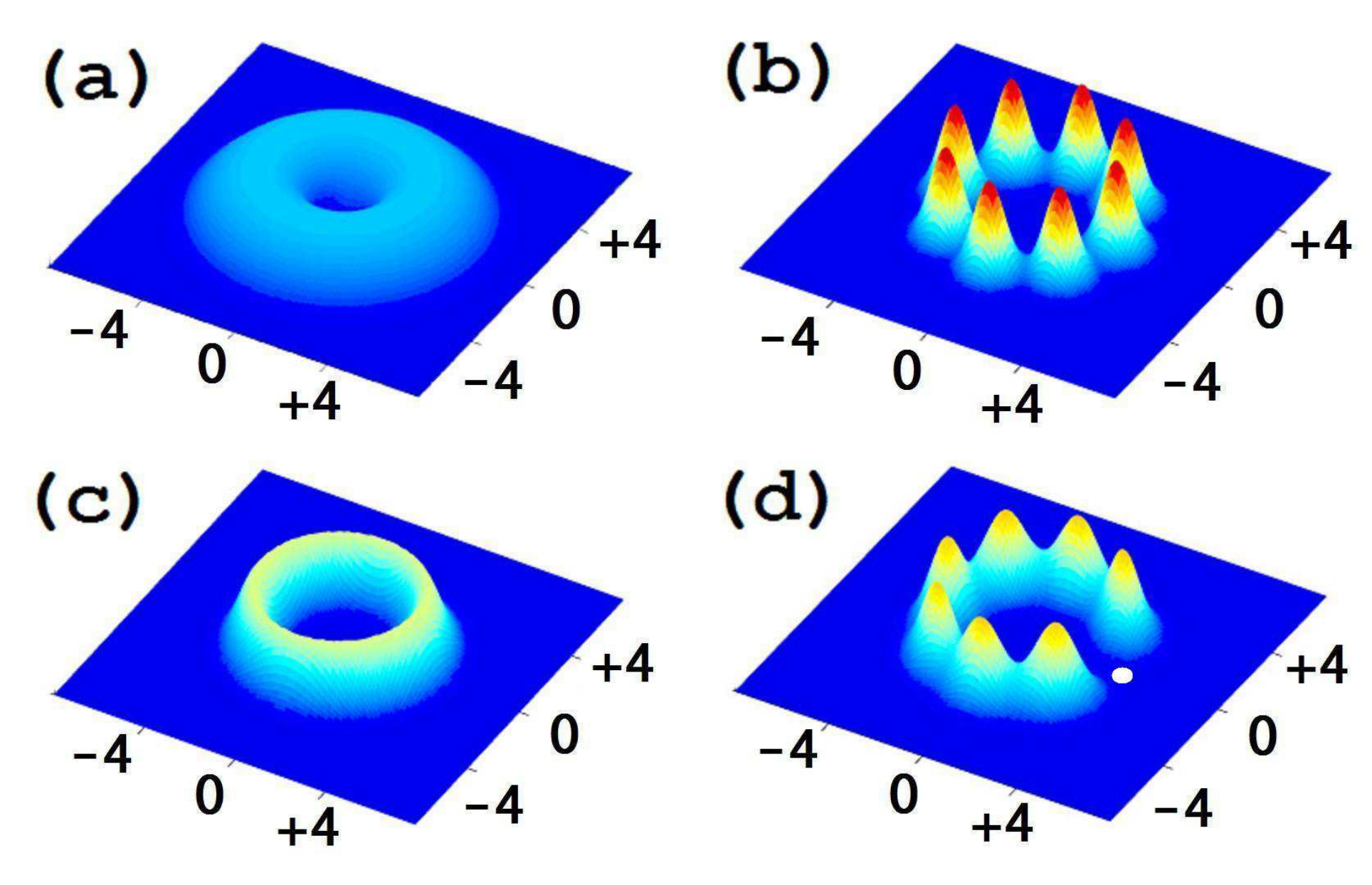}
\caption{
Single-particle densities and CPDs for $N=8$ neutral repelling bosons in a rotating 2D toroidal trap
with reduced rotational frequency $\Omega/\omega_0=0.2$ and $R_\delta=50$. 
The confining potential, $m \omega_0^2 (r-r_0)^2/2$, is centered at a radius $r_0=3 l_0$. 
(a) Gross-Pitaevski SPD. (b) UHF SPD exhibiting
breaking of the circular symmetry. (c) QWM SPD exhibiting circular
symmetry. (d) CPD for the QWM wave function (resulting from the method
of symmetry restoration), revealing the hidden point-group symmetry in the
intrinsic frame of reference. The fixed observation point is denoted by a white dot.
The QWM ground-state angular momentum is $L_z=16$. Lengths in units of the oscillator 
length $l_0$. The vertical scale is the same for (b), (c), and (d), but
different for (a).
}
\end{figure}

\begin{figure}[t]
\centering\includegraphics[width=5.50cm]{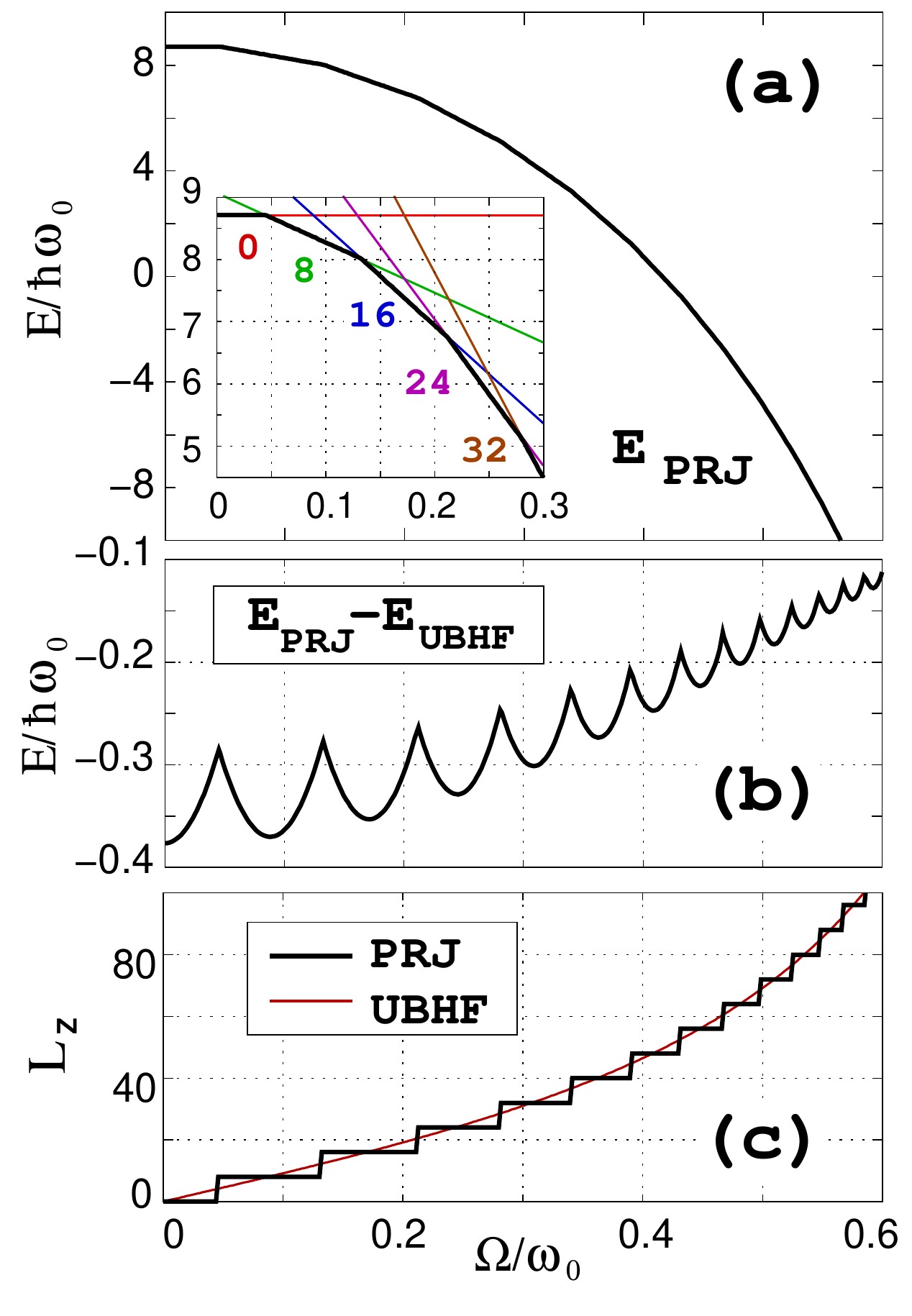}
\caption{
Properties of $N=8$ neutral repelling bosons in a rotating 2D toroidal 
trap as a function of the reduced rotational frequency $\Omega/\omega_0$.  
The confining toroidal potential is centered at a radius $r_0=3 l_0$, and the 
interaction-strength parameter was chosen as $R_\delta=50$. 
(a) QWM ground-state energies, $E^{PRJ}$. The term QWM corresponds to projected (PRJ) 
wave functions that preserve the total angular momentum (symmetry restoration). 
The inset shows the range $0 \leq \Omega/\omega_0 \leq 0.3$. The numbers denote
ground-state magic angular momenta. 
(b) Energy difference $E^{PRJ}-E^{UBHF}$, where the subscript ``UBHF'' stands for
unrestricted bosonic Hartree-Fock. 
(c) Total angular momenta associated  with (i) the QWM ground states [thick solid
line (showing steps and marked as PRJ); online black] and (ii) 
the broken-symmetry UBHF solutions (smooth thin solid line; online red).
}
\end{figure}

Of great value in analyzing the physics associated with the hidden symmetries is the evolution of the 
lowest-energy band in the energy spectra (yrast band) as a function of the two successive approximations 
(broken symmetry UHF $\rightarrow$ symmetry restoration). Figs.\ 3(a,b) present the evolution of yrast 
spectra (as a function of the rotational frequency $\Omega$ of the trap) that are associated with the class
of wave functions portrayed in Fig.\ 2. The most prominent trend is that the ground-state angular momenta 
of the symmetry-restored wave functions do not assume all the possible 2D values, but are restricted 
to stepwise values $L_z=N k$, $k=0,1,2,\ldots$, with $N=8$, i.e., they change in steps of $N=8$, where $N$ 
is the  number of particles [see Fig.\ 3(c) and the inset in Fig.\ 3(a)]. Such stepwise angular momenta are
usually referred to as ``magic'' and the associated ground states of enhanced stability [see Fig.\ 3(b)] 
are finite-size precursors of the bulk fractional quantum Hall states; see also Section 3 below.  

\section{Group theoretical analysis of symmetry breaking in unrestricted Hartree Fock}

\begin{figure}[t]
\centering\includegraphics[width=6.50cm]{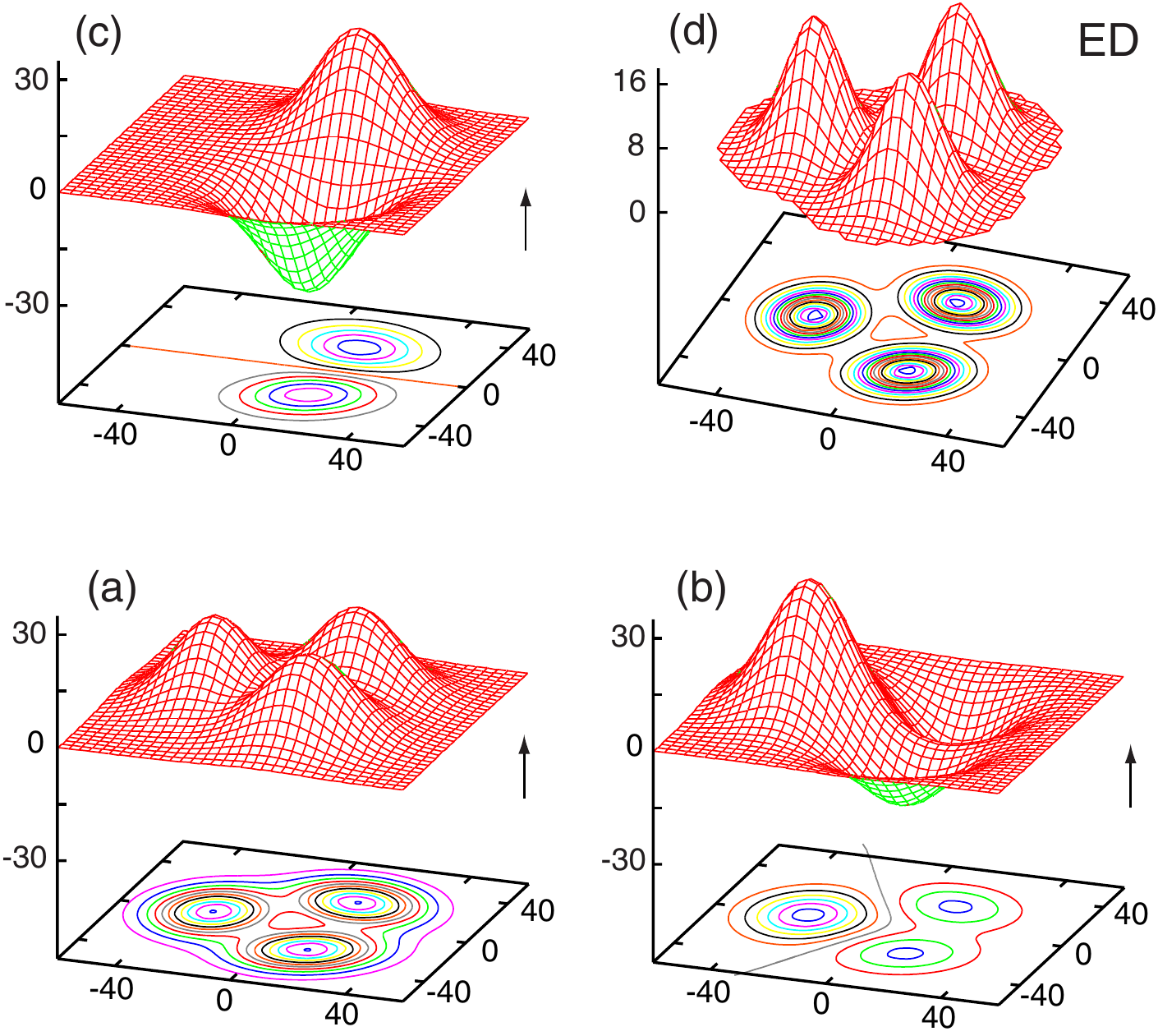}\\
~~~~\\
\caption{
The UHF case exhibiting breaking of the circular symmetry for $N=3$ electrons and total
spin projection $S_z=3/2$ at $R_W=10$ and at zero magnetic field. (a-c): real orbitals (modulus square). 
(d): the corresponding electron density (ED).
The choice of the remaining parameters is: $\hbar \omega_0=5$ meV 
and effective mass $m^*=0.067 m_e$.
Distances are in nanometers. The real orbitals are in
10$^{-3}$ nm$^{-1}$ and the total ED in 10$^{-4}$ nm$^{-2}$.
The arrows indicate the spin projection ($S_z=1/2$) for each orbital.
}
\end{figure}

We mention here the case of $N=3$ fully spin polarized $(S_z=3/2)$ electrons
in the absence of a magnetic field ($B$) and for $R_W=10$ $(\kappa=1.9095)$. Fully
spin polarized UHF determinants preserve the total spin, but for this
value of $R_W$ the lowest in energy UHF solution is one with broken circular 
symmetry. As it has been mentioned earlier, broken rotational symmetry does not imply 
no space symmetry, but a lower point-group symmetry \cite{yann03}. 

In Fig.\ 4 we display the UHF symmetry-violating orbitals (a$-$c) whose 
energies are (a) 44.801 meV, (b) and (c) 46.546 meV, namely the two orbitals 
(b) and (c) with the higher energies are degenerate in energy. 
Overall the BS orbitals (a$-$c) drastically differ from the orbitals 
of the independent particle model. In particular, they are associated with 
specific sites (within the QD) forming an equilateral triangle, and thus they 
can be described as having the structure of a linear combination of ``atomic'' 
(site) orbitals (LCAOs). Such LCAO molecular orbitals (MOs) are familiar in 
natural molecules, and this analogy supports the term ``semi-classical electron (or Wigner) 
molecules'' for characterizing the BS-UHF solutions. 
We notice here that the LCAO orbitals in Fig.\ 4 are familiar in Organic Chemistry and 
are associated with the theoretical description of Carbocyclic Systems, and in
particular the molecule C$_3$H$_3$ (cyclopropenyl, see, e.g., Ref.\ 
\cite{cot}). The important point of course is not the uniqueness or not
of the 2D UHF orbitals, but the fact that they transform according to the
irreducible representations of specific point groups, leaving both the
UHF determinant and the associated electron densities invariant.

The electron density (ED) portrayed in Fig.\ 4(d) remains invariant under certain 
geometrical symmetry operations, namely those of an unmarked, 
plane and equilateral triangle. They are: (1) The identity $E$; (2) The two 
rotations $C_3$ (rotation by $2\pi/3$) and $C_3^2$ (rotation by $4 \pi/3$);
and (III) The three reflections $\sigma_v^I$, $\sigma_v^{II}$, and 
$\sigma_v^{III}$ through the three vertical planes, one passing through each
vertex of the triangle. These symmetry operations for the unmarked equilateral
triangle constitute the elements of the group $C_{3v}$ \cite{cot,wol}.
 
One of the main applications of group theory in Chemistry is the determination
of the eigenfunctions of the Schr\"{o}dinger equation by simply using symmetry 
arguments alone. This is achieved by constructing the 
so-called {\it symmetry-adapted linear combinations\/} (SALCs) of AOs. 
A widely used tool for constructing SALCs is the projection operator
\begin{equation}
\hat{{\cal P}}^\mu = \frac{n_\mu}{|{\cal G}|} \sum_R \chi^\mu(R) \hat{R},
\label{prch}
\end{equation}
where $\hat{R}$ stands for any one of the symmetry operations of the molecule,
and $\chi^{\mu}(R)$ are the characters of the 
$\mu$th irreducible representation
of the set of $\hat{R}$'s. (The $\chi^\mu$'s are tabulated in the socalled
character tables \cite{cot,wol}.) $|{\cal G}|$ denotes the order of the group
and $n_\mu$ the dimension of the representation.

The task of finding the SALCs for a set of three $1s$-type AOs exhibiting 
the $C_{3v}$ symmetry of an equilateral triangle can be simplified, since the 
pure rotational symmetry by itself (the rotations $C_3$ 
and $C_3^2$, and not the reflections $\sigma_v$'s through the 
vertical planes) is sufficient for their determination. Thus one needs to 
consider the simpler character table of the cyclic group $C_3$ 
(see Table 1).
\begin{table}[t]
\caption{Character table for the cyclic group $C_3$ [$\varepsilon = 
\exp(2\pi i/3)]$}
\begin{center}
\begin{tabular}{c|ccc}
$C_3$ & $E$ & $C_3$ & $C_3^2$ \\ \hline
~~$A$~~ & 1 & 1 & 1 \\
~~$E^\prime$~~ & 1 & $\varepsilon$ & $\varepsilon^*$ \\
~~$E^{\prime\prime}$~~ & 1 & $\varepsilon^*$ & $\varepsilon$ \\
\end{tabular}
\end{center}
\end{table}

From Table 1, one sees that the set of the three $1s$ AOs situated 
at the vertices of an equilateral triangle spans the two irreducible
representations $A$ and $E$, the latter one consisting of two associted
one-dimensional representations. To construct the SALCs, one simply
applies the three projection operators $\hat{{\cal P}}^A$, 
$\hat{{\cal P}}^{E^\prime}$, and $\hat{{\cal P}}^{E^{\prime\prime}}$ to
one of the original AOs, let's say the $\phi_1$,
\begin{eqnarray}
\hat{{\cal P}}^A \phi_1 & \approx &
(1) \hat{E} \phi_1 + (1) \hat{C}_3 \phi_1 + (1) \hat{C}_3^2 \phi_1 
 =  (1) \phi_1 + (1) \phi_2 + (1) \phi_3 \nonumber \\
& = & \phi_1 + \phi_2 + \phi_3,
\label{paf1}
\end{eqnarray}
\begin{equation}
\hat{{\cal P}}^{E^\prime} \phi_1  \approx 
(1) \hat{E} \phi_1 + (\varepsilon) \hat{C}_3 \phi_1 + 
(\varepsilon^*) \hat{C}_3^2 \phi_1 
 =  \phi_1 + \varepsilon \phi_2 + \varepsilon^* \phi_3,
\label{paf2}
\end{equation}
\begin{equation}
\hat{{\cal P}}^{E^{\prime\prime}} \phi_1  \approx 
(1) \hat{E} \phi_1 + (\varepsilon^*) \hat{C}_3 \phi_1 + 
(\varepsilon) \hat{C}_3^2 \phi_1 
 = \phi_1 + \varepsilon^* \phi_2 + \varepsilon \phi_3.
\label{paf3}
\end{equation}
The $A$ SALC in Eq.\ (\ref{paf1}) is real. The two $E$ SALCs [Eq.\ (\ref{paf2}) and 
Eq.\ (\ref{paf3})], however, are complex functions and do not coincide with 
the real UHF orbitals. These complex SALCs agree with 
BS-UHF orbitals obtained in the case of an applied magnetic field. 
On the other hand, a set of two real and orthogonal SALCs 
that spans the $E$ representation can be derived fron Eq.\ (\ref{paf2}) and 
Eq.\ (\ref{paf3}) by simply adding and substracting the two complex ones. This
procedure recovers immediately the real UHF orbitals displayed in Fig.\ 4. 

\section{Restoration of circular symmetry: Group structure and sequences of magic angular momenta}

In the previous section, we discussed how the BS-UHF determinants and 
orbitals describe indeed 2D molecular stuctures (semi-classical Wigner molecules) 
in close analogy with the case of natural 3D molecules. However, the study of 
the WMs at the UHF level restricts their description to the {\it intrinsic\/}
(nonrotating) frame of reference. Motivated by the case of natural atoms, one 
can take a subsequent step and address the properties of 
{\it collectively\/} rotating QWMs in the 
laboratory frame of reference. As is well known, for natural atoms, this step 
is achieved by writing the total wave function of the molecule as the product
of the electronic and ionic partial wave functions. In the case of the purely
electronic or bosonic WMs, however, such a product wave function requires the assumption
of complete decoupling between intrinsic and collective degrees of freedom,
an assumption that might be justifiable in limiting cases only.

Using the BS UHF solutions, this subsequent step can be addressed by using the 
post-Hartree-Fock method of {\it restoration of broken symmetries\/} \cite{yann07,rs} 
via projection (PRJ) techniques. 

In this section, we will use the PRJ approach to illustrate 
how certain universal properties of the CI (exact) solutions, 
i.e., the appearance of magic angular momenta in the exact rotational spectra, 
\cite{yann07,gir,mc,rua,sek,mak} relate to the symmetry broken UHF solutions. 
Indeed, we will demonstrate that the magic angular momenta are a direct
consequence of the symmetry breaking at the UHF level and that they are 
determined fully by the molecular symmetries of the UHF 
determinant.

As an illustrative example, we have chosen the relatively simple, but non 
trivial case, of $N=3$ electrons. For $B=0$, both the $S_z=1/2$ and $S_z=3/2$
polarizations can be considered. We start with the $S_z=1/2$ polarization,
whose BS UHF solution (let's denote it by $|\downarrow \uparrow \uparrow 
\rangle$) exhibits a breaking of the total spin symmetry in addition to the rotational symmetry.
We first proceed with the restoration of the total spin by noticing that
$|\downarrow \uparrow \uparrow \rangle$ has a point-group symmetry lower 
than the $C_{3v}$ symmetry of an equilateral triangle. 
The $C_{3v}$ symmetry, however, can be readily restored by applying the 
projection operator in Eq.\ (\ref{prch}) to $|\downarrow \uparrow \uparrow \rangle$ 
and by using the character table of the cyclic $C_3$ group (see Table 1). Then 
for the intrinsic part of the many-body wave function, one finds two different 
three-determinantal combinations, namely
\begin{equation}
\Phi_{intr}^{E^\prime} (\gamma_0)= 
|\downarrow \uparrow \uparrow \rangle
+ e^{2\pi i/3} |\uparrow \downarrow \uparrow \rangle
+ e^{-2\pi i/3} |\uparrow \uparrow \downarrow \rangle,
\label{3dete1}
\end{equation}
and
\begin{equation}
\Phi_{intr}^{E^{\prime\prime}} (\gamma_0)=
|\downarrow \uparrow \uparrow \rangle
+ e^{-2\pi i/3} |\uparrow \downarrow \uparrow \rangle
+ e^{2\pi i/3} |\uparrow \uparrow \downarrow \rangle,
\label{3dete2}
\end{equation}
where $\gamma_0=0$ denotes the azimuthal angle of the vertex associated with
the original spin-down orbital in $|\downarrow \uparrow \uparrow \rangle$.
We note that the intrinsic wave functions $\Phi_{intr}^{E^\prime}$
and $\Phi_{intr}^{E^{\prime\prime}}$ are eigenstates of the square of
the total spin operator ${\hat{\bf S}}^2$  ($\hat{\bf S} = \sum_{i=1}^3
\hat{\bf s}_i$) with quantum number $s(s+1)=3/4$, $(s=1/2)$). This can be verified directly by 
applying ${\hat {\bf S}}^2$ to them.

To restore the circular symmetry in the case of a $(0,N)$ ring arrangement, one 
applies the projection operator \cite{yann07,rs,yann02},
\begin{equation}
2 \pi {\cal P}_I \equiv \int_0^{2 \pi}
d\gamma \exp[-i \gamma (\hat{L}-I)]~,
\label{amp}
\end{equation}
where $\hat{L}=\sum_{j=1}^N \hat{l}_j$ is the operator for the total angular
momentum. Notice that the operator ${\cal P}_I$ is a direct generalization of
the projection operator in Eq.\ (\ref{prch}) to the case of the continuous cyclic
group $C_\infty$ [the phases $\exp(i \gamma I)$ are the characters of 
$C_\infty$].

The projected wave function, $\Psi_{PRJ}$, (having both good total 
spin and angular momentum quantum numbers) is of the form,
\begin{equation}
2 \pi \Psi_{PRJ} = \int^{2\pi}_0 d\gamma
\Phi_{intr}^E (\gamma) e^{i\gamma I},
\label{rbsi}
\end{equation}
where now the intrinsic wave function [given by Eq.\ (\ref{3dete1}) or
Eq.\ (\ref{3dete2})] has an arbitrary azimuthal orientation $\gamma$. We note 
that, unlike the phenomenological Eckardt-frame model \cite{mak} where 
only a single product term is involved, the PRJ wave function in Eq.\ 
(\ref{rbsi}) is an average over all azimuthal directions of an infinite
set of product terms. These terms are formed by multiplying the UHF intrinsic 
part $\Phi_{intr}^{E}(\gamma)$ with the external rotational wave 
function  $\exp(i \gamma I)$ (the latter is properly
characterized as ``external'', since it is an eigenfunction of the total
angular momentum $\hat{L}$ and depends exclusively on the azimuthal
coordinate $\gamma$).

The operator ${\hat{R}(2\pi/3) \equiv \exp (-i 2\pi{\hat L}/3})$ can be
applied onto $\Psi_{{PRJ}}$ in two different ways, namely either on
the intrinsic part $\Phi_{{intr}}^{E}$ or the external part $\exp(i \gamma
I)$. Using Eq.\ (\ref{3dete1}) and the property $\hat{R}(2\pi/3) 
\Phi_{{intr}}^{E^\prime} =\exp (-2\pi i/3)\Phi_{{intr}}^{E^\prime}$, 
one finds,
\begin{equation}
\hat{R}(2\pi/3) \Psi_{{PRJ}} = \exp (-2\pi i/3) \Psi_{{PRJ}},
\label{r1rbs}
\end{equation}
from the first alternative, and
\begin{equation}
\hat{R}(2\pi/3) \Psi_{{PRJ}} = \exp (-2\pi I i/3) \Psi_{{PRJ}},
\label{r2rbs}
\end{equation}
from the second alternative. Now if $\Psi_{{PRJ}} \neq 0$, the only
way that Eqs.\ (\ref{r1rbs}) and (\ref{r2rbs}) can be simultaneously true is
if the condition $\exp [2\pi (I-1) i/3]=1$ is fulfilled. This leads to a first
sequence of magic angular momenta associated with total spin $s=1/2$, i.e.,
\begin{equation}
I = 3 k +1,\; k=0,\pm 1, \pm 2, \pm 3,...
\label{i1}
\end{equation}

Using Eq.\ (\ref{3dete2}) for the intrinsic wave function, and following
similar steps, one can derive a second sequence of magic angular momenta
associated with good total spin $s=1/2$, i.e.,
\begin{equation}
I = 3 k -1,\; k=0,\pm 1, \pm 2, \pm 3,...
\label{i2}
\end{equation}

In the fully polarized case, the UHF determinant 
is denoted as  $|\uparrow \uparrow \uparrow\;\rangle$, 
and it is already an eigenstate of $\hat{\bf S}^2$ with quantum number 
$s=3/2$. Thus only the rotational symmetry needs to be restored, that is,
the intrinsic wave function is simply $\Phi^A_{{intr}}(\gamma_0) = 
|\uparrow \uparrow \uparrow \;\rangle$. Since 
$\hat{R}(2\pi/3) \Phi^A_{{intr}} = \Phi^A_{{intr}}$, the condition
for the allowed angular momenta is $\exp [-2\pi I i/3]=1$, which yields
the following magic angular momenta,
\begin{equation}
I = 3 k,\; k=0,\pm 1, \pm 2, \pm 3,...
\label{i3}
\end{equation}

We note that in high magnetic fields only the fully polarized case is
relevant and that only angular momenta with $k > 0$ enter in Eq.\ (\ref{i3})
(see Refs.\ \cite{yann02.2,yann03.2}). In this case, in the thermodynamic limit, the 
partial sequence with $k=2q+1$, $q=0,1,2,3,...$ is directly related to the odd 
filling factors $\nu=1/(2q+1)$ of the fractional quantum Hall effect
[via the relation $\nu = N(N-1)/(2I)$]. This suggests that the observed 
hierarchy of fractional filling factors in the quantum Hall effect may be 
viewed as a signature originating from the point group symmetries of the 
intrinsic wave function $\Phi_{{intr}}$, and thus it is a manifestation
of symmetry breaking at the UHF mean-field level. 

\section{Summary}

The analysis presented above concerning the relation between hidden symmetries
and emergent signatures (e.g., magic amgular momenta) in the spectra of
symmetry-restored mean-field wave functions applies also in the case of 
configuration-interaction, exact many body wave functions; see the review in Ref.\ \cite{yann07}.
The CI method requires large-scale, parallel computations, but it has the advantage
of providing benchmark results due to the achieved high quantitative accuracy.
We refer to this combination of mean-field and CI analysis as computational
microscopy \cite{yann15}. 

Recently, we have incorporated in our CI computer codes the option of Dirac-delta
contact interactions, in addition to the long-range Coulomb one.
Thus we have been able to analyze the wave function anatomy 
of a few untracold fermionic ($^6$Li) atoms in single and double optical traps,
where the formation of quantum Wigner molecules can be associated with the
emergence of Heisenberg antiferromagnetic behavior \cite{yann15,yann09,murm15} and the 
creation of highly entangled states; e.g., the celebrated Bell states for two $^6$Li atoms.
(Such behavior was earlier predicted in the case of strongly repelling electrons in 
double quantum dots \cite{yann09}.) 
The unprecedented experimental control of the interparticle interaction (from zero to 
infinite strength) achieved in the case of a few trapped ultracold atoms is enabling 
investigations of fundamental physics (such as high T$_c$ and 1D and 2D magnetism) from 
a bottom-up perspective. In addition, in analogy with the field of 2D semiconductor double 
and triple QDs, it promises technological applications in the area of quantum information
and quantum computing. Our computational-microscopy approach can be used to investigate
the universal behavior in the strongly correlated regime of both ultracold fermionic or
bosonic trapped atoms and confined electrons. 

\section*{Acknowledgements}

This work is supported by the Office of Basic Energy
Sciences of the US Department of Energy (Grant No. FG05-
86ER45234).

%
%

\end{document}